\documentclass[journal]{IEEEtran}

\newif\iflong
\longtrue

\usepackage[english]{babel}

\usepackage{ifpdf}

\iflong\else

\IEEEaftertitletext{\vspace{-2\baselineskip}}
\fi

\usepackage{cite} 
\usepackage{url}
\usepackage{hyperref}

\ifCLASSINFOpdf
	\usepackage[pdftex]{graphicx}
	\graphicspath{{./figures/}}
\else
	\usepackage[dvips]{graphicx}
	\graphicspath{./figures/}
\fi
\usepackage{color}
\usepackage[table]{xcolor} 

\usepackage{pgf, tikz, pgfplots, circuitikz}
\pgfplotsset{compat=1.16}
\usetikzlibrary{shapes, arrows, automata, plotmarks}
\usetikzlibrary{calc, hobby, decorations}

\usepackage[cmex10]{amsmath} 
\usepackage{amsfonts, amssymb, amsthm}
\usepackage{mathrsfs}

\usepackage{algorithm,algpseudocode,algorithmicx}
    \algnewcommand{\LeftComment}[1]{\Statex \(\triangleright\) #1}

\usepackage{enumerate}
\usepackage{multirow}
\usepackage{rotating}
\usepackage[caption=false,font=normalsize,labelfont=sf,textfont=sf]{subfig}

\usepackage{booktabs}

\usepackage{needspace}

\input{auxFiles/mySymbol.sty}
\input{auxFiles/pennColors.sty}

\usepackage{subfiles}

\begin{document}

\title{Unsupervised Optimal Power Flow \\ Using Graph Neural Networks}

\date{}

\author{Damian~Owerko,~
	Fer\hspace{0.015cm}nando~Gama,~
	and~Alejandro~Ribeiro
	\iflong
		\thanks{Supported by NSF CCF 1717120, ARO W911NF1710438, ARL DCIST CRA W911NF-17-2-0181, ISTC-WAS and Intel DevCloud. D. Owerko and A. Ribeiro are with the Dept. of Electrical and Systems Eng., Univ. of Pennsylvania., F. Gama is with the Electrical and Computer Engineering Department, Rice University, Houston, TX.  Email: \{owerko,aribeiro\}@seas.upenn.edu, and fgama@rice.edu.
		}
	\fi
}


\maketitle


\begin{abstract}
	Optimal power flow (OPF) is a critical optimization problem that allocates power to the generators in order to satisfy the demand at a minimum cost. Solving this problem exactly is computationally infeasible in the general case. In this work, we propose to leverage graph signal processing and machine learning. More specifically, we use a graph neural network to learn a nonlinear parametrization between the power demanded and the corresponding allocation. We learn the solution in an unsupervised manner, minimizing the cost directly. In order to take into account the electrical constraints of the grid, we propose a novel barrier method that is differentiable and works on initially infeasible points. We show through simulations that the use of GNNs in this unsupervised learning context leads to solutions comparable to standard solvers while being computationally efficient and avoiding constraint violations most of the time.
\end{abstract}


\IEEEpeerreviewmaketitle


\section{Introduction} \label{sec:intro}



Optimal power flow (OPF) is a critical optimization problem for the energy industry.
It consists in allocating power to each generator in the grid, so that the energy demand is satisfied with minimum cost (optimally).
OPF is used to allocate electricity generation throughout the day, establish day-ahead market prices, and plan for grid infrastructure \cite{Cain12-ACOPF-History}.


The objective of the OPF problem is to minimize the cost of generating electrical power, subject to the constraints imposed by the grid infrastructure, the physical laws of electromagnetism, and the demand patterns.
This problem is nonconvex due to the sinusoidal nature of the alternating current and voltage and the constraints imposed by electrical interconnections of the grid.
Therefore, solving the OPF problem exactly is computationally infeasible in the general case \cite{Cain12-ACOPF-History,nonconvex}.
In fact, it has been show to be NP-hard \cite{bienstock}.


One of the most common ways used to address the non-tractability of the OPF problem is to solve a linear surrogate based on small-angle approximations. \cite{Sun10-DCOPF}.
In practical cases, however, power grids are typically highly loaded thus violating the small-angle approximation \cite{Chatzivasileiadis18-Lecture}.
A more accurate, but computationally intensive solution, is to rely on solvers for interior point methods \cite{acopf-computational}.
The OPF problem can also be approximated by nonlinear, convex optimization problems.
Examples include quadratic \cite{Hijazi2017-QuadraticRelaxation, Sundar2018-StrengthenedQuadraticRelaxation}, second order conic \cite{Jabr2006-SecondOrderConic} and semi-definite programming (SDP) \cite{Bai2007-SDP, Jabr2012-SDP} relaxations. There exists a small class of topologies for which sufficient conditions exist for convex relaxation optimality \cite{Low14-ConvexRelaxationExactness}. In particular, these relaxations work well for radial networks, but often lead to infeasible solutions or sub-optimal for meshed topologies \cite{Low14-ConvexRelaxationExactness}.
Nevertheless, SDP relaxations were shown to perform well on a variety of topologies \cite{Molzahn2016-ConvexRelaxationStrength}, promting research into more advanced relaxations with more general optimality guarantees \cite{Kocuk2016-StrongSOCPRelaxations}.
An alternative approach is to use successive linear approximations that converge to the exact formulation \cite{Castillo2016-IVACOPF}. This allows to balance trade-offs between accuracy and computation time and integrates well into existing infrastructure that relies on LP solvers. The implementation is slower than interior point methods. However, unlike interior point methods, successive linear approximations can be implemented on LP solvers already used in industry.

Machine learning has arisen as a promising approach to overcome computational tractability.
Inference of machine learning models is typically much faster than traditional solvers and feasibility of the solution can be quickly verified or used to hot-start a computationally expensive solver. Historically, there were many attempts to apply machine learning to this problem, with the work in \cite{AlRashidi2009-ComputationalIntelligence} providing a comprehensive review until 2009. Many of these approaches such as genetic algorithms and particle swarms, do scale well and therefore have not between shown to be effective for networks with more than 30 buses \cite{Bakirtzis2002-Optimal, Todorovski2003-Power, Wang2005-modified, Pham22-ReducedOPF-GNN}.
Early approaches used imitation learning to learn to replicate solutions obtained using interior point methods \cite{gnn_opf_imitation, guha}, but often violated constraints and could not perform that the method used to train them. Newer approaches such as DeepOPF \cite{Pan2021-DeepOPF}, use constrained learning and are able to provide more robust solutions.

Graph signal processing (GSP) has emerged as a convenient mathematical framework to describe problems involving network data \cite{Shuman13-SPG, Ortega18-GSP}.
By extending concepts of traditional signal processing, such as filtering and frequency analysis, to graph-based data, GSP provides novel tools for the analysis and design of distributed solutions \cite{Segarra17-Linear, Sandryhaila13-DSPG}.
Of particular interest are graph neural networks (GNNs) \cite{Gama20-GNN}, which have been shown to be successful in both signal processing and machine learning problems involving graph data \cite{Gama19-Stability, Eisen19-Wireless, Tolstaya19-Flocking}.
GNNs are built as extensions of graph convolutional filters, followed by (typically pointwise, typically nonlinear) activation functions.
This allows GNNs to learn nonlinear behaviors while retaining a decentralized nature and exploiting the underlying graph structure.
Thus, GNNs are promising candidates for learning optimal power flow allocations from state measurements, while respecting the topology of the grid.

In this paper we propose a novel, unsupervised, constrained learning approach. We make the following contributions.
\begin{enumerate}
    \item We use GNNs to parametrize a mapping between the state of the buses in the grid and the target generated power.
    \item We learn the resulting parametrization by means of optimizing the constrained OPF problem.
    \item We introduce a differentiable, piece-wise penalty function based on the log-barrier in order to enforce constraints. This allows for the use of gradient-based methods.
    \item We propose new methods of evaluating machine learning OPF models. Since we are dealing with a physical system, it is important to evaluate not only the rate at which a model violates constraints, but also the severity of those violations.
    \item We show that graph neural networks are ideally suited to optimal power flow as they scale well for sparse graphs and can be implemented in a distributed manner.
\end{enumerate}

Section \ref{sec:opf} introduces the OPF problem. Section \ref{sec:GNN} leverages graph signal processing to present an appropriate description of the OPF problem and introduces the GNN-based parametrization of the solution. In section \ref{sec:unsupervisedOPF} we propose a novel approach to unsupervised learning of OPF, where piece-wise penalty functions \ref{subsec:penalty_choice} are used as a method of enforcing constraints \ref{subsec:penalty_choice}. Finally, we provide experimental results \ref{sec:sims} on the efficacy of our architecture on the IEEE 30 and 118 bus power system test cases.

\section{Optimal Power Flow} \label{sec:opf}



An electrical grid is an interconnected system that generates electricity and delivers it to consumers.
Its two most important electrical components are generators that produce electricity and loads that demand electric power \cite{allen_textbook}.
Denote by $\ccalB^{G}$ be the set of all generators such that generator $i \in \ccalB^G$ produces $S^g_i \in \mbC$ units of power.
Similarly, load $i \in \ccalB^{D}$ demands $S^d_i \in \mbC$ units of power.
Note that the power is described by means of a complex number to indicate its active (real) and reactive (imaginary) power.
The cost $c_{i}$ to operate a generator is a function $c_i : \mbC \mapsto \mbR$ of the power produced.

The problem of OPF is concerned with minimizing the generation costs, while meeting the demand and satisfying the electrical constraints of the grid \cite{Chatzivasileiadis18-Lecture}. Specifically, in OPF we are trying to find the generator output power $S_{i}^{g}$ which minimize the total production cost
\begin{equation}
    \min_{\{S_i^g\}_{i \in \ccalB^G}} \sum_i c_i(S_i^g). \label{eq:cost}
\end{equation}
The quantity of power produced by each generator is constrained within a certain range by two complex limits
\begin{equation}
    S^g_{i,\min} \preceq S^g_i \preceq S^g_{i,\max} \label{eq:generator_limit}
\end{equation}
for all $i \in \ccalB^{G}$, and where $S^g_{i,\min},S^g_{i,\max} \in \mathbb{C}$ and $\preceq$ is a generalized inequality over the complex plane.
The lower bound for the generator output is often zero, but not necessarily.
Some generators, like nuclear power plants, cannot be turned off within the time-frame of the optimization problem \cite{Lokhov2011-LoadFollowing}.
Other generators might be able to store power leading to negative lower bounds.

A \emph{bus} is a grid element to which all other electrical components connect. Consider the set of all buses $\ccalB$. Each generator and load is connected to exactly one bus. Denote the sets of generators and loads connecting to bus $i \in \ccalB$ as $\ccalB^G_i \subseteq \ccalB^G$ and $\ccalB^D_i \subseteq \ccalB^D$, respectively. There may be multiple generators and loads connected to one bus.

Additionally, the bus $i \in \ccalB$ is characterized by its voltage $V_i \in \mbC$, the net power injected at the bus $S_i \in \mbC$, and its shunt admittance $Y^s_i$. The shunt admittance is the combined admittance between a bus' connected devices and ground. Therefore, the net power injected is the total power injected by generators on the bus minus the power consumed by the load on the bus, which can be conveniently written as follows. \cite{powermodels}
\begin{equation}
    S_i = \sum_{j \in \ccalB^G_i} S^g_j - \sum_{j \in \ccalB^D_i} S^d_j \label{eq:bus_power}
\end{equation}
The power lost due to the shunt admittance is given by Ohm's law $Y_i^s|V_i|^2$ \cite{powermodels}.

The physical limitations of the components connected to a bus constrain the voltage magnitude between $V_{i,\min}, V_{i,\max} \in \mbR$ \cite{powermodels}. That is,
\begin{equation}
    V_{i,\min} \le |V_i| \le V_{i,\max}. \label{eq:voltage_magnitude}
\end{equation}
While the components may be rated to operate at voltages outside this range, eq. \eqref{eq:voltage_magnitude} defines the stable operating conditions \cite{powermodels}.

\begin{figure}
    \centering
    \begin{tikzpicture}[european]
	\draw 
		(0,0) node[above]{$V_i$} to[oosourcetrans, *-, a={$T_{ij}:1$}]
		++(3,0) coordinate (A) to[generic, l={$Y_{ij}$}]
		++(3,0) coordinate (B) to[short, -*]
		++(0.5,0) node[above]{$V_j$};
		
	\draw (A) to[generic, *-, l={$Y^c_{ij}$}] ++(0,-1.5) node[tground]{};
	\draw (B) to[generic, *-, l={$Y^c_{ij}$}] ++(0,-1.5) node[tground]{};
\end{tikzpicture}
    \caption{Circuit diagram of $\pi$-section branch model with transformer.}
    \label{fig:branch}
\end{figure}
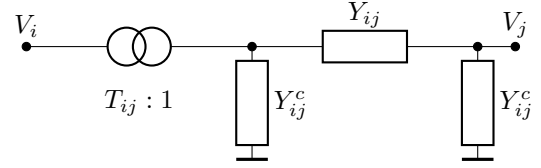

\emph{Branches} connect different buses with each other. A branch $(i,j) \in \ccalE \subseteq \ccalB \times \ccalB$ connects bus $i$ to bus $j$. We model branches using an ideal transformer and a line in series, a model known as a $\pi$-section branch model with a transformer, see figure \ref{fig:branch} for a circuit diagram and \cite{matpower, powermodels, PowerModels18-ModelVideo} for further details. An ideal transformer is characterized by the complex transformation ratio $T_{ij} \in \mbC$ which describes the ratio of the input voltage to the output voltage \cite{matpower,powermodels}, i.e. $V_{j} = V_i/T_{ij}$, and assumes no internal losses. This transformer is in series with a $\pi$-section line, which is described by three parameters: the line admittance $Y_{ij} \in \mbC$ describing the flow of current from one end to the other, the forward charging admittance $Y^c_{ij} \in \mbC$ and backward charging admittance $Y^c_{ji} \in \mbC$. Note that the air is a dielectric and therefore we model its effect on the line by adding a capacitor connected to the ground. Therefore, the total power flowing forward $S_{ij} \in\mbC$ and backward $S_{ji} \in\mbC$ along a branch is \cite{powermodels}
\begin{align}
    S_{ij} & = (Y_{ij} + Y^c_{ij})^* \frac{|V_i|^2}{|T_{ij}|^2} - Y_{ij}^* \frac{V_iV_j^*}{T_{ij}} \label{eq:branch_forward} \\
    S_{ji} & = (Y_{ij} + Y^c_{ji})^* |V_j|^2 - Y_{ij}^* \frac{V_i^*V_j}{T_{ij}^*}. \label{eq:branch_backward}
\end{align}
There is a physical limit to the amount power a branch can handle for extended periods of time \cite{matpower,powermodels}. This can be represented by $S_{ij,\max} \in \mbR$ such that
\begin{align}
    |S_{ij}| & \le S_{ij,\max}. \label{eq:power_max}
\end{align}
Additionally a branch may have limits, $\Delta\theta_{ij,\min}$ and $\Delta\theta_{ij,\max}$ in $\mbR$, on the difference in voltage phase angle between the buses it connects \cite{powermodels}. Specifically,
\begin{equation}
    \Delta\theta_{ij,\min} \le \angle(V_iV_j^*) \le \Delta\theta_{ij,\max} \label{eq:angle_difference}
\end{equation}
where $\angle(V_{i}V_{j}^{\ast})$ is the angle difference between the voltage phase at bus $i$ and the voltage phase at bus $j$. It is common to define reference buses to eliminate ambiguity in terms of voltage angles \cite{matpower, powermodels}, which are defined to have zero voltage angle.

Buses and branches are related by the power-flow equation as a direct consequence of Kirchhoff's current law \cite{powermodels}. The net power injected at the bus \eqref{eq:bus_power} must be equal to the power flowing out of the bus, so it holds that
\begin{align}
    S_i = Y_i^s|V_i|^2 - & \sum_{(i,j) \in \ccalE \cap \ccalE^T} S_{ij} \label{eq:power_flow}.
\end{align}

The distinction between the power flow and the optimal power flow problems is important \cite{matpower, Chatzivasileiadis18-Lecture}. In the \emph{power flow problem} the goal is to solve for the voltage $V_i$, given the net power injected $S_i$ at each node and the topological characteristics of the grid, as described by the values $Y_{i}^{s}, T_{ij}, Y_{ij}, Y_{ij}^{c}$. The power generated and demanded at each node are exogenous variables. \emph{Optimal power flow} (OPF), on the other hand, is a constrained optimization problem where the power generated is endogenous. The goal is to determine the power output of each generator $S_i^g$ that minimizes the total generation cost $\sum_{i \in \ccalB} c_i(S_i^g)$ while satisfying power flow equation \eqref{eq:power_flow} and the aforementioned constraints. In this problem, only the power demanded at each node $S_i^d$ is an exogenous variable. Table \ref{tab:OPF} summarizes the optimal power-flow problem.

\begin{table}
    \caption{The complete optimal power flow equations for the formulation used in this paper. Refer to section \ref{sec:opf} for explanations of individual equations.}
    \label{tab:OPF}
    \hrule
    \begin{align}
        \textbf{minimize}\quad   &
        \sum_{i \in \ccalB^G} c_i(S^g_i) \tag{\ref{eq:cost}}                                                                                                                                 \\
        \textbf{subject to}\quad & \nonumber                                                                                                                                                 \\
                                 & S^{g}_{i,\min} \le S^g_i \le S^{g}_{i,\max}, \;\forall\; i \in \ccalB^G \tag{\ref{eq:generator_limit}}                                                    \\
                                 & S_i = \sum_{j \in \ccalB^G_i} S^g_i - \sum_{j \in \ccalB^D_i} S^d_i. \tag{\ref{eq:bus_power}}                                                             \\
                                 & V_{i,\min} \le |V_i| \le V_{i,\max}, \;\forall\; i \in \ccalB  \tag{\ref{eq:voltage_magnitude}}                                                           \\
                                 & S_{ij} = (Y_{ij} + Y^c_{ij})^* \frac{|V_i|^2}{|T_{ij}|^2} - Y_{ij}^* \frac{V_iV_j^*}{T_{ij}}, \;\forall\; (i,j) \in \ccalE  \tag{\ref{eq:branch_forward}} \\
                                 & S_{ji} = (Y_{ij} + Y^c_{ji})^* |V_j|^2 - Y_{ij}^* \frac{V_i^*V_j}{T_{ij}^*}, \;\forall\; (i,j) \in \ccalE  \tag{\ref{eq:branch_backward}}                 \\
                                 & |S_{ij}| \le S_{ij,\max}, \;\forall\; (i,j) \in \ccalE  \tag{\ref{eq:power_max}}                                                                          \\
                                 & \Delta\theta_{ij,\min} \le \angle(V_i V_j^*) \le \Delta\theta_{ij,\max}, \;\forall\; (i,j) \in \ccalE  \tag{\ref{eq:angle_difference}}                    \\
                                 & S_i = Y^s_i|V_i|^2 + \sum_{(i,j) \in \ccalE \cup \ccalE^T} S_{ij}, \;\forall\; i \in \ccalB  \tag{\ref{eq:power_flow}}
    \end{align}
    \hrule
\end{table}

\subsection{Solutions}\label{subsec:solutions}

There is extensive research into solving the optimal power flow problem \cite{Cain12-ACOPF-History}. The branch equations,  \eqref{eq:branch_forward} and \eqref{eq:branch_backward}, make it non-convex \cite{nonconvex}, as the complex multiplications involve trigonometric functions. Finding the exact solution is strongly NP-hard \cite{bienstock}. Consequently, many research efforts focus on finding approximations, relaxations, and solutions for OPF on special families of graphs.

One common approach, named DC-OPF, is a linear approximation to the exact OPF problem (sometimes referred to as AC-OPF). DC-OPF is a first order approximation around the point where voltage angles are close to zero. Since the voltage angle differences are small, complex multiplications in \eqref{eq:branch_forward} and \eqref{eq:branch_backward} are approximated using small-angle approximations. The problem becomes linear if we normalize voltage magnitude. Additionally, if the cost function is convex, so too is the DC-OPF problem. Most commonly, the cost function is a second-order polynomial. Nevertheless, DC-OPF solutions are not guaranteed to be feasible in the exact OPF case \cite{Wang21-FeasibilityACDC}. Particularly, the assumptions of the DC-OPF problem are violated when demand is high, which coincidentally is the most critical case of the grid operation \cite{Wang21-FeasibilityACDC, Chatzivasileiadis18-Lecture}.

In some situations, the OPF problem can be solved exactly by using using one of several off-the-shelf optimization problem solvers, such as CONOPT, IPOPT, KNITRO, MINOS or SNOPT \cite{Castillo13-Survey}. However, in general, they are slow to converge for large networks \cite{acopf-computational} or have no guarantee of convergence. In particular, the IPOPT (Interior Point OPTimizer) \cite{ipopt} solver has found widespread use due to its robustness, but it is computationally costly \cite{acopf-computational}.


\section{Graph Neural Networks} \label{sec:GNN}



The objective of this work is to approximate the OPF solution by taking the power demanded by each load $S^d_i$ for all  $i \in \ccalB^D$ as an input and outputting the optimal generation scheme, $S^g_i$ for all $i \in \ccalB^G$, in a scalable and distributed way. To do so, we parametrize the solution by means of a graph neural network.

\subsection{Graph Signal Processing} \label{subsec:GSP}

Graph signal processing (GSP) has emerged as a convenient framework to describe, analyze and solve problems that are distributed in nature. Let $\ccalG = (\ccalB, \ccalE, \ccalW)$ be a graph where $\ccalB = \{1,...,N\}$ is the set of $N$ nodes, $\ccalE \subseteq \ccalB \times \ccalB$ is the set of edges, and $\ccalW: \ccalE \to \reals$ is a weight function that assigns a (positive) scalar to each edge. Data is described as a signal $z: \ccalB \to \reals^F$ defined on top of the nodes of the graph. The signal at a node is a vector of $F$ measurements or \textit{features}. It is often convenient to represent a graph signal as a matrix $\bbZ \in \mbR^{N \times F}$. The $i$th row of $\bbZ$, $\bbz_i \in \reals^{F}$, collects the $F$ features at node $i$, $\bbz_i = z(i)$ \cite{Sandryhaila13-DSPG, Shuman13-SPG, Ortega18-GSP}. We represent complex quantities by a pair of features.

Representing graph signals as a $N \times F$ matrix, while convenient, does not capture the underlying graph support present in the definition $z : \ccalB \to \reals^{F}$. To address this, a real symmetric matrix $\bbA \in \reals^{N \times N}$, known as a \emph{graph shift operator} (GSO), is employed. The matrix $\bbA$ is a GSO if its elements satisfy $[\bbA]_{ij} = 0$ whenever $(j,i) \notin \ccalE$. Most commonly used are the graph adjacency matrix, the Laplacian, and their normalized versions \cite{Gama20-GNN, Ortega18-GSP}. Multiplying a graph signal by a GSO produces a \emph{shifted} version of the signal. This is a local operation, i.e. each component of the shifted signal linearly combines information from neighboring nodes. If the neighborhood of node $i \in \ccalB$ is $\ccalN(i) = \{ j \in \ccalB : (j,i) \in \ccalE \}$ then
\begin{equation}
    [\bbA \bbZ]_{if} = \sum_{j=1}^N [\bbA]_{ij} [\bbZ]_{jf} = \sum_{j \in \ccalN(i) \cup i} [\bbA]_{ij} [\bbZ]_{jf}. \label{eq:shift_local}
\end{equation}
Note that the first equality is simply the definition of matrix multiplication, while the second arises from the sparsity pattern of the GSO $\bbA$.

GSOs are the building blocks for convolutions on graphs. In essence, the GSO is a generalization of the unit time-shift operator from traditional signal processing. Continuing with the analogy, a graph convolution is a linear shift-and-sum operation, whereby a graph signal $\bbZ$ is weighed by a sequence of $K+1$ \emph{filter taps}, $\bbH_k \in \reals^{F \times F'}$ for $k=0,...,K$
\begin{equation}
    H(\bbZ; \bbA) = \sum_{k=0}^K \bbA^k \bbZ \bbH_k. \label{eq:graph_filter}
\end{equation}
and where $F$ and $F'$ is the number of input and output features, respectively. Equation \eqref{eq:graph_filter} represents the graph convolution implementation of a linear shift-invariant graph filter $H: \reals^{N \times F} \mapsto \reals^{N \times F'}$ which is a linear map between two graph signals with different feature sizes. To draw further analogies with filtering, we observe that the output of the graph filter \eqref{eq:graph_filter} is the result of applying a bank of $FF'$ linear shift-invariant graph filters \cite{Segarra17-Linear}.

Note that the notion of locality for the shift operator extends to graph filtering. As analyzed above, the shift operation \eqref{eq:shift_local} involves only one-hop communication to produce a linear combination of adjacent signal values. Likewise, repeated application of $\bbA$ computes a linear combination of values located farther away. That is, $\bbA^{k} \bbZ$ collects the feature values at nodes in the $k$-hop neighborhood. The value of $\bbA^{k} \bbZ = \bbA (\bbA^{k-1} \bbZ)$ can be computed locally by $k$ repeated exchanges with the one-hop neighborhood. We note that multiplication by $\bbH_{k}$ on the right does not affect the locality of the graph filter \eqref{eq:graph_filter}, since $\bbH_{k}$ only mixes features available at each single node. That is, it takes the $F$ input features, and mixes them linearly to obtain $F'$ new features. If each node stores the filter taps $\{\bbH_k\}$ it can compute the output of the filter locally, communicating only with its $K$-hop neighborhood. Therefore, graph filtering is both a local and a distributed linear processing architecture.

\subsection{Graph Neural Networks}\label{subsec:gnn}

\iflong
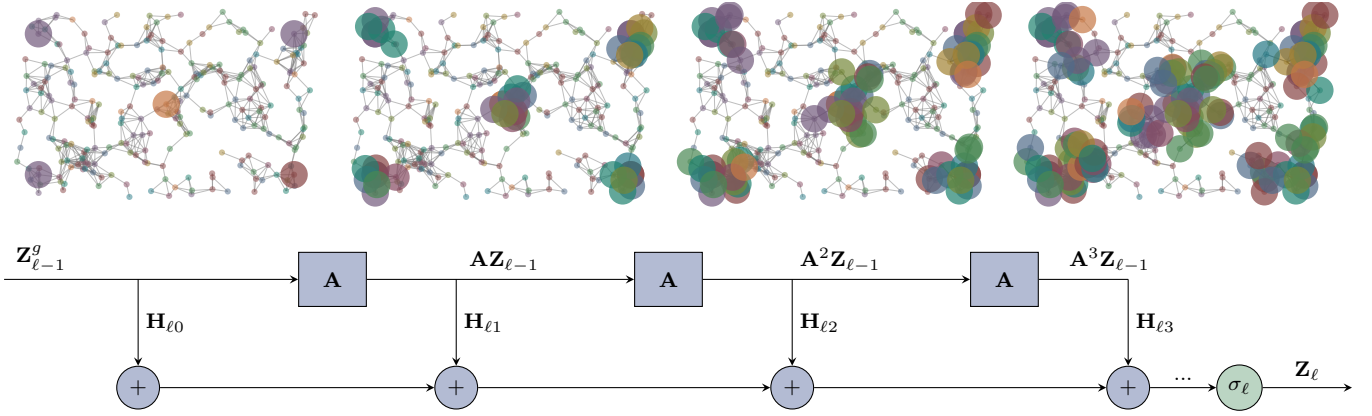
\begin{figure*}[t]
    \centering
    \includegraphics [width = 0.24\linewidth]
    {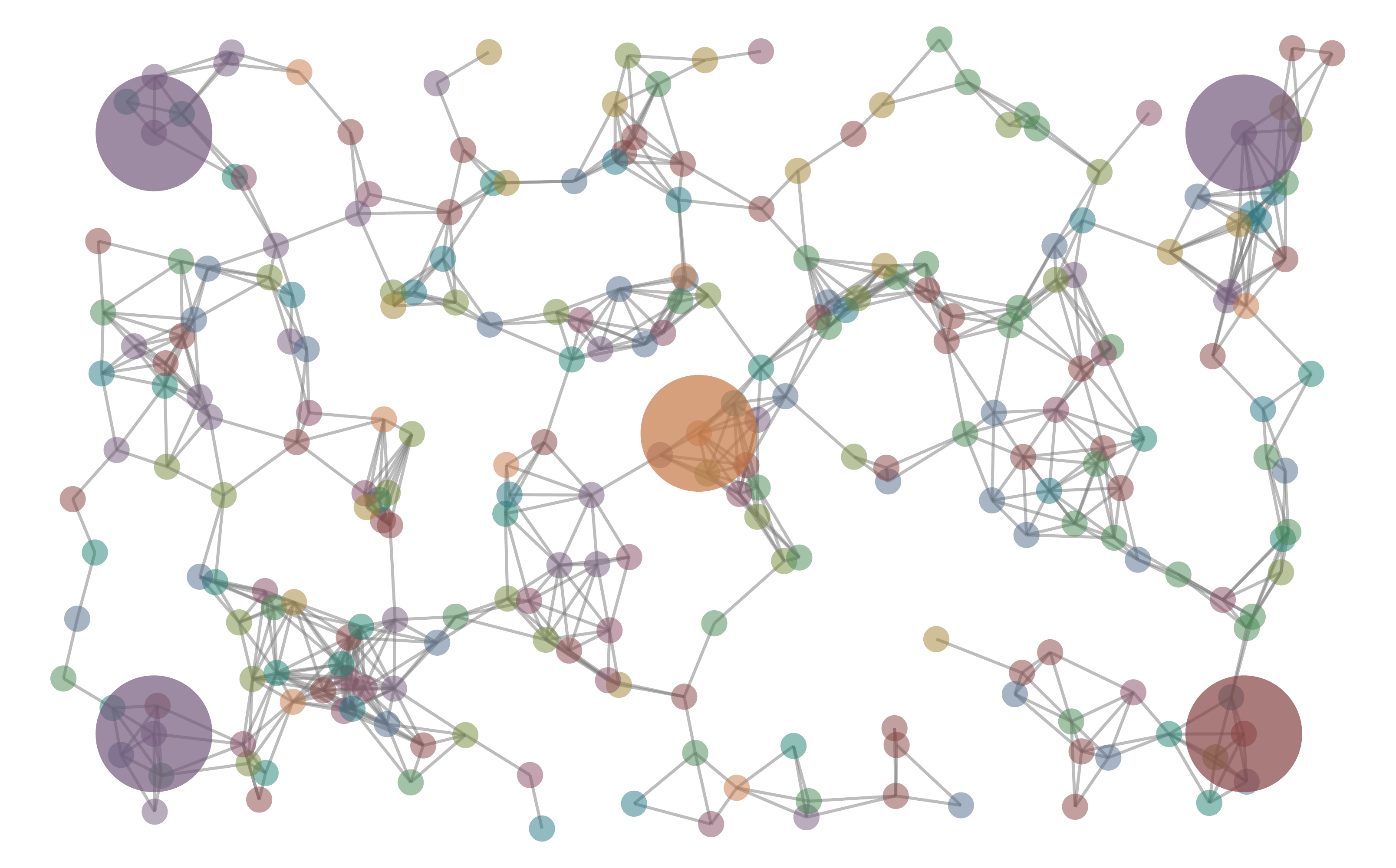}
    \includegraphics [width = 0.24\linewidth]
    {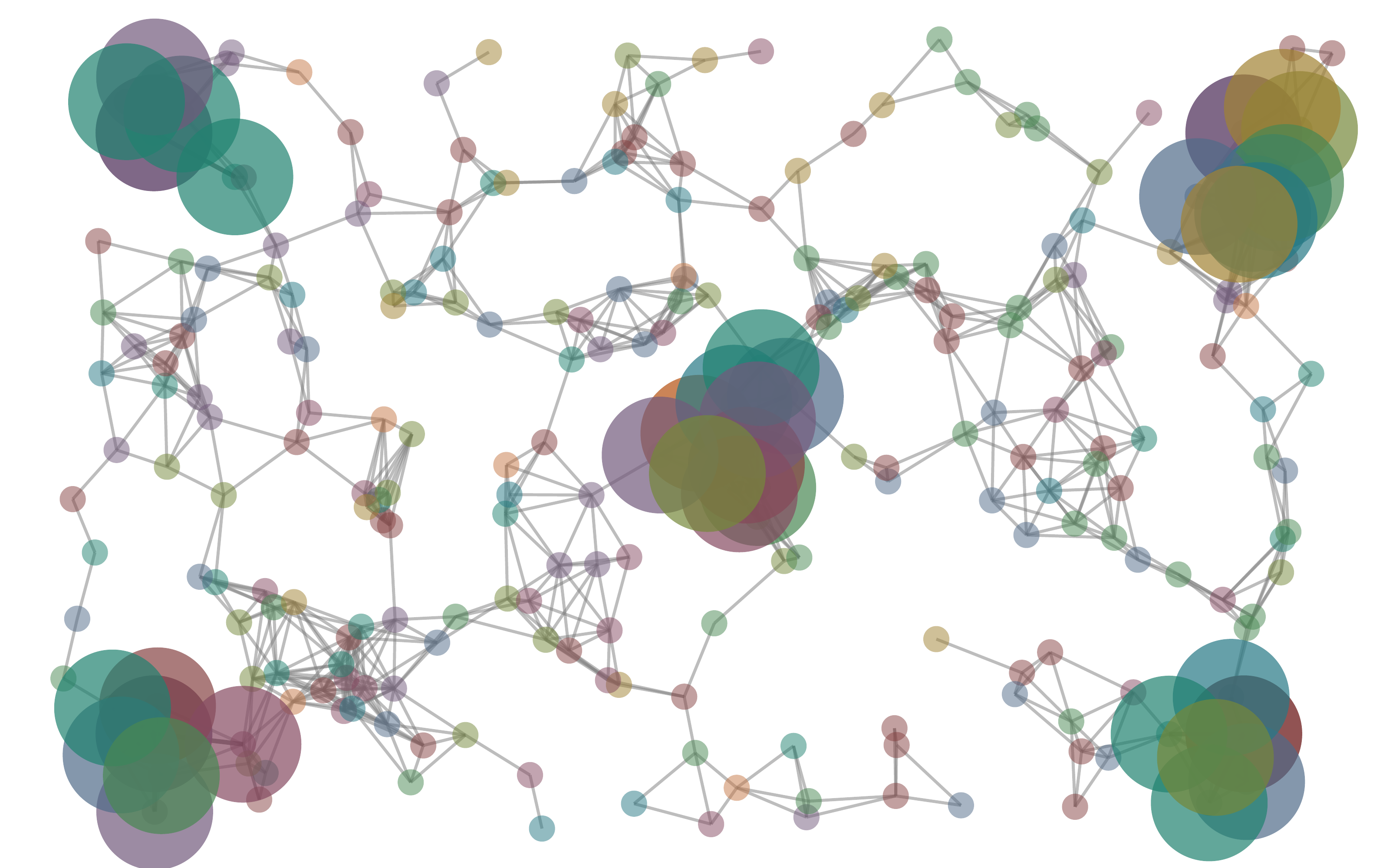}
    \includegraphics [width = 0.24\linewidth]
    {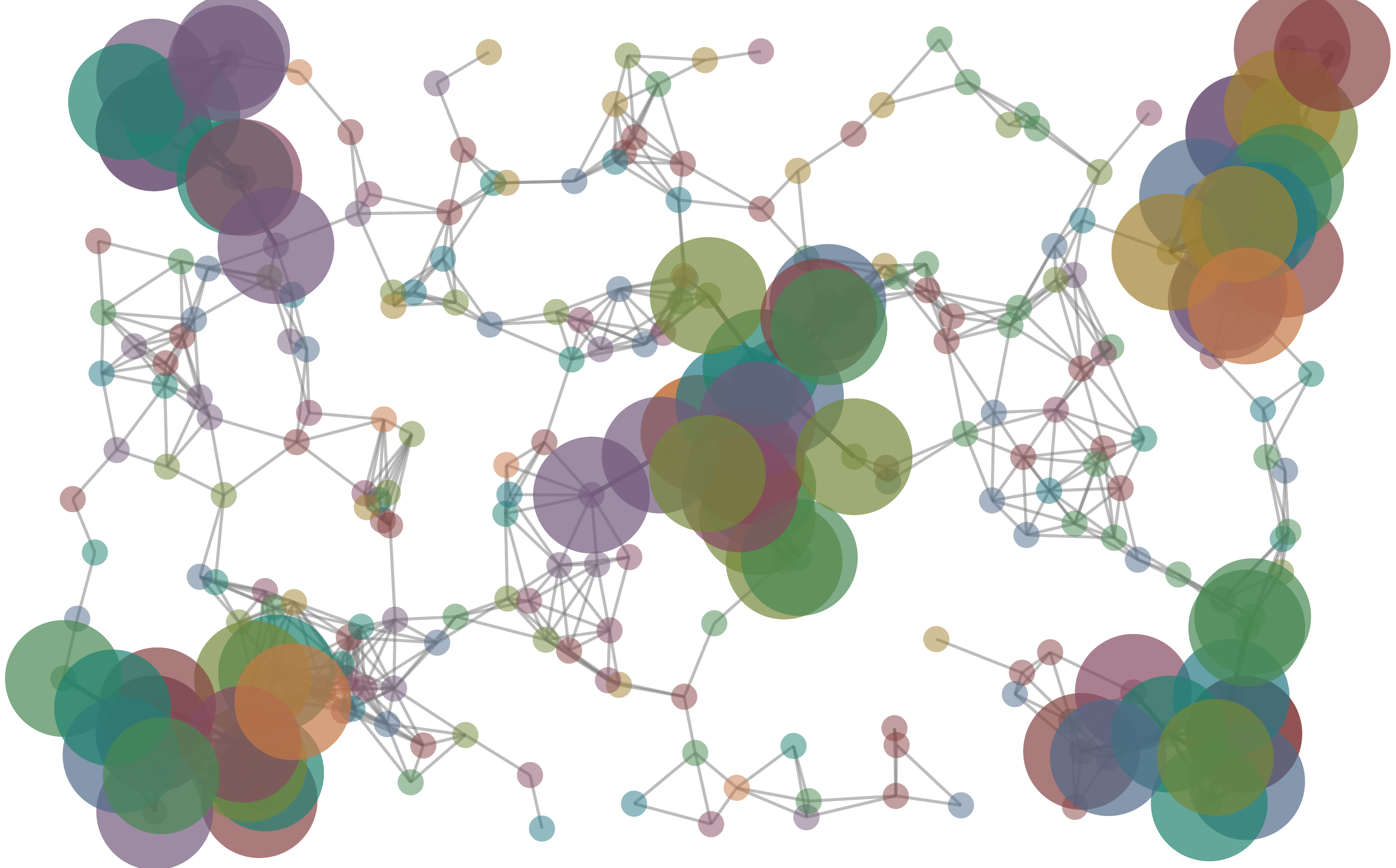}
    \includegraphics [width = 0.24\linewidth]
    {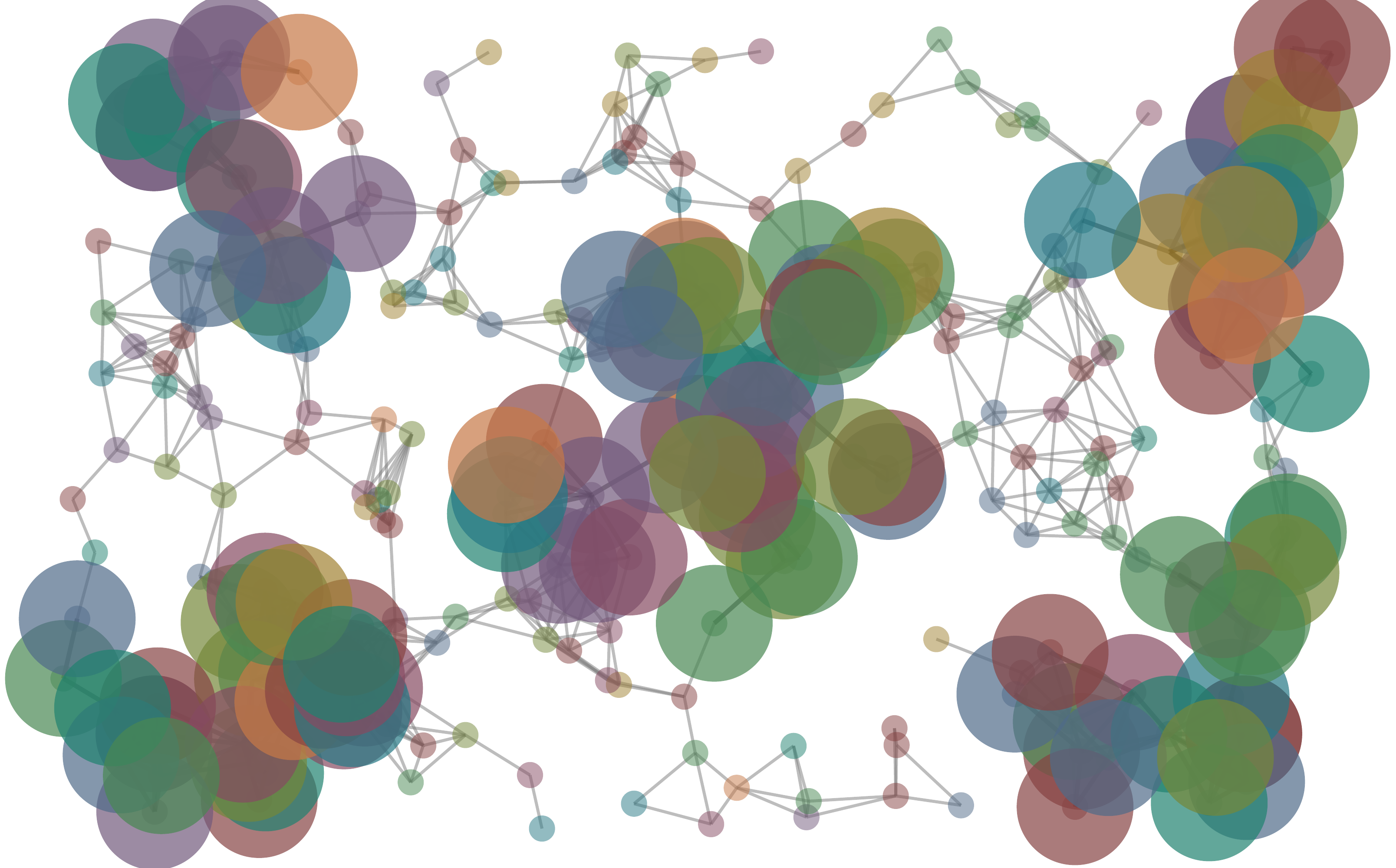} \\ \bigskip

\def \thisplotscale {1.8}
\def \unit {\thisplotscale cm}

\tikzstyle {Phi} = [rectangle, 
                    thin,
                    minimum width = 0.5*\unit, 
                    minimum height = \sumshift*\unit, 
                    anchor = west,
                    draw,
                    fill = pennblue!20]

\tikzstyle {sum} = [circle, 
                    thin,
                    minimum width  = 0.3*\unit, 
                    minimum height = 0.3*\unit, 
                    anchor = center,
                    draw,
                    fill = pennblue!20]

\def \deltax {1.65}
\def \deltay {0.8}
\def \sumshift {0.4}

\begin{tikzpicture}[x = 1*\unit, y = 1*\unit] \footnotesize
    
\node (origin) [] {};
\path (origin) ++ (0*\deltax, 0) node (first) [] {};
    
\path (first) ++ (1.35*\deltax, 0) node (0) [Phi] {$\bbA$};
\path (0)     ++ (1.35*\deltax, 0) node (1) [Phi] {$\bbA$};
\path (1)     ++ (1.35*\deltax, 0) node (2) [Phi] {$\bbA$};

\path (2.east) ++ (1.0*\sumshift*\deltax, 0) node [anchor=west] (last) [] {};

\path (first.east) ++ (1.5*\sumshift*\deltax, -\deltay) node (sum0) [sum] {$+$};
\path (0.east) ++ (\sumshift*\deltax, -\deltay) node (sum1) [sum] {$+$};
\path (1.east) ++ (\sumshift*\deltax, -\deltay) node (sum2) [sum] {$+$};
\path (2.east) ++ (\sumshift*\deltax, -\deltay) node (sum3) [sum] {$+$};

\path[-stealth] (first) edge [very near start, above] node {$\bbZ_{\ell-1}^{g}$}               (0);	
\path[-stealth] (0)     edge [above] node {$\ \bbA\bbZ_{\ell-1}$}     (1);	
\path[-stealth] (1)     edge [above] node {$\ \bbA^{2}\bbZ_{\ell-1}$} (2);	
\path[-]        (2)     edge [above, near end] node {$\ \bbA^{3}\bbZ_{\ell-1}$} (sum3 |- last);

\path[-stealth, draw] (sum0 |- first) -- (sum0) node [midway, right] {$\bbH_{\ell 0}$};	
\path[-stealth, draw] (sum1 |- 0)     -- (sum1) node [midway, right] {$\bbH_{\ell 1}$};	
\path[-stealth, draw] (sum2 |- 1)     -- (sum2) node [midway, right] {$\bbH_{\ell 2}$};	
\path[-stealth, draw] (sum3 |- 2)     -- (sum3) node [midway, right] {$\bbH_{\ell 3}$};

\path[-stealth, draw] (sum0) -- (sum1);	
\path[-stealth, draw] (sum1) -- (sum2);	
\path[-stealth, draw] (sum2) -- (sum3);	

\path (sum3.east) ++ (0.4 * \deltax, 0) node (nonlinearity) [sum, fill=penngreen!20] {$\sigma_{\ell}$};
\path[-stealth,draw] (sum3) -- (nonlinearity) node [midway, above] {$\displaystyle ...$};

\path[-stealth] (nonlinearity.east) edge [midway, above] node {$\bbZ_{\ell}$} ++ (0.4*\deltax, 0);



\end{tikzpicture}
    \caption{Graph neural networks. Every node takes its data value $\bbZ_{\ell-1}$ and weighs it by $\bbH_{\ell 0}$ (first graph). Then, all the nodes exchange information with their one-hop neighbors to build $\bbA\bbZ_{\ell-1}$, and weigh the result by $\bbH_{\ell 1}$ (second graph). Next, they exchange their values of $\bbA\bbZ_{\ell-1}$ again to build $\bbA^{2}\bbZ_{\ell-1}$ and weigh it by $\bbH_{\ell 2}$ (third graph). This procedure continues for $K+1$ steps until all $\bbA^{k} \bbZ_{\ell-1}\bbH_{\ell k}$ have been computed for $k=0,\ldots,K$, and added up to obtain the output of the graph convolution operation \eqref{eq:graph_filter}. Then, the non-linearity $\sigma_{\ell}$ is applied to compute $\bbZ_{\ell}$. To avoid cluttering, this operation is illustrated on only $5$ nodes. In each case, the corresponding neighbors accessed by successive relays of information are indicated by the colored disks.} 
    \label{fig:selection}
\end{figure*}
\fi

A graph neural network (GNN) is a nonlinear map $\bbPhi(\bbZ; \ccalH, \bbA)$ that is applied to the input $\bbZ$ and takes into account the underlying graph $\ccalG$ via a GSO $\bbA$. A GNN consists of a cascade of $L$ layers, each of them a graph filter \eqref{eq:graph_filter} followed by a point-wise non-linearity $\sigma_{\ell}$ \iflong(see Fig.~\ref{fig:selection} for an illustration)\fi,
\begin{equation} \label{eq:GNN}
    \bbPhi(\bbZ;\ccalH, \bbA) = \bbZ_{L},\: \bbZ_{\ell} = \sigma_{\ell} [H_\ell(\bbZ_{\ell-1}; \bbA)]
\end{equation}
for $\ell=1,\ldots,L$, where $\bbZ_{0}=\bbU$ is the input signal \cite{Bruna14-DeepSpectralNetworks, Defferrard17-CNNGraphs, Gama19-Architectures}. The output $\bbZ_{\ell} \in \reals^{N \times F_{\ell}}$ of layer $\ell$ is a graph signal with $F_{\ell}$ features and the output of the last layer $\bbZ_{L}$ is the output of the GNN. The specific non-linearity $\sigma_\ell$, the number of features $F_\ell$ and the number of filter weights $K_\ell$ are design choices. The filter weights $\ccalH = \{ \bbH_{k\ell} \in \reals^{F_{\ell-1} \times F_\ell}, k = 0,\ldots,K_\ell, \ell = 0,...,L \}$ are model parameters to be learned from data by a training process. Notice that GNNs are a straightforward generalization of convolutional neural networks (CNNs), where traditional convolutions are replaced with graph convolutions. Alternatively, GNNs are a simple nonlinear extension of graph convolutions \cite{Gama19-Architectures}.

The computation of the intermediate output $\bbZ_{\ell}$ in each of the $\ell$ layers can be carried out entirely in a local and distributed manner, through repeated exchanges with one-hop neighbors. The total number of parameters in $\ccalH$ is $\sum_{\ell=1}^{L} F_{\ell-1}F_{\ell} K_{\ell}$, independent of the size $N$ of the graph. Thus, the GNN \eqref{eq:GNN} is a scalable architecture \cite{Gama19-Architectures}. This justifies GNNs as the model of choice for electrical grids which are networks with thousands of nodes and sparse connectivity.

Furthermore, GNNs exhibit the properties of permutation equivariance and stability to graph perturbations \cite{Gama19-Stability}. The former allows the GNN to learn from fewer datapoints by exploiting the topological symmetries of the graph. The latter allows the GNN to have a good performance when used on different graphs than trained on, as long as these graphs are similar. Importantly, the GSO is an input to the GNN, $\bbPhi(\bbZ; \ccalH, \bbA)$. Therefore, the parameters, $\ccalH$, learned from training on one graph, can be reused for inference on another.


\section{Unsupervised Learning for OPF} \label{sec:unsupervisedOPF}




We are interested in solving the OPF problem by learning a solution that acts as a mapping between the demanded power $\bbS_{i}^{d}$ for all $i \in \ccalB^{D}$ and the required generated power $\bbS_{i}^{g}$ for all $i \in \ccalB^{G}$. To do this, we propose to parametrize the mapping by means of a GNN, thus leveraging their local and distributed properties, as well as their scalability and transferability. In Section~\ref{subsec:OPF-GSP} we describe the OPF problem in the framework of GSP, while in Section~\ref{subsec:penalty_method} we describe the learning process as solving a constrained optimization problem. In Section~\ref{subsec:penalty_choice} we discuss the choice of penalty functions for improved convergence.

\subsection{Optimal power flow as a graph signal processing problem} \label{subsec:OPF-GSP}

Since the OPF is a problem distributed on the topology of the power grid, we can leverage GSP to describe it and find novel solutions. Recall from section \ref{sec:opf} that $\ccalB$ is the set of buses and $\ccalE$ is the set of branches. Hence, let $\ccalG$ be a weighted graph with nodes $\ccalB$ and edges $\bar{\ccalE} \subseteq \ccalE$. It has edge weights $\ccalW(i,j) := w_{ij}$, which depend on the admittance $Y_{ij}$ between bus $i$ and $j$. We will define the weights as the Gaussian kernel 
\begin{equation}\label{eq:graph_weights}
    w_{ij} := \exp(-\alpha/|Y_{ij}|^2)
\end{equation}
where $\alpha$ is a scaling factor. We ignore branches whose weight is less than a threshold $\beta$, so that $\bar{\ccalE} := \{ (i,j) \in \ccalE \mid w_{ij} > \beta \}$.
The values of $\alpha$ and $\beta$ are hyperparameters, adequately chosen as discussed in Section~\ref{sec:sims}. Denote by $\bbA \in \reals^{N \times N}$ the adjacency matrix of $\ccalG$ such that the elements of $\bbA$, $[\bbA]_{ij} = w_{ij}$ if $(i,j) \in \bar{\ccalE}$ and $0$ otherwise.
Since the graph is undirected, the matrix $\bbA$ is symmetric. 
By construction, it is a GSO on $\ccalG$. 
For simplicity, we ignore the line charging admittance $Y_{ij}^c$, assuming that it is negligible \cite{matpower}.

In day-to-day grid operation the network parameters of the electrical grid (see section \ref{sec:opf}), such as branch impedance $Y_{ij}$ and voltage limits $V_{i,\min}$ and $ V_{i,\max}$ are known \emph{a priori}. Therefore, the \emph{state} of the buses is fully described by the power injected, $S_i$ and complex voltage, $V_i$, at each node. These two complex quantities can be described by four real graph signals, $\bbp, \bbq, \bbv, \bbdelta \in \reals^N$, namely real power injections, complex power injections, voltage magnitude and voltage angle.
\begin{subequations}\label{eq:state_vectors}
\begin{align}
    \bbp &= \bmat{\Re{S_1} & \dots & \Re{S_N}}^T\\
    \bbq &= \bmat{\Im{S_1} & \dots & \Im{S_N}}^T\\
    \bbv &= \bmat{|V_1| & \dots & |V_N|}^T\\
    \bbdelta &= \bmat{\angle V_1 & \dots & \angle V_N}^T
\end{align}
\end{subequations}
Alternatively, the bus state can be thought of as a single graph signal with multiple features, $\bbX \in \reals^{N \times 4}$. It is a concatenation of the state vectors from \eqref{eq:state_vectors}.
\begin{equation}\label{eq:state}
    \bbX = \begin{bmatrix} \bbp & \bbq & \bbv & \bbdelta \end{bmatrix}
\end{equation}
Notice that the bus state fully determines the branch state through Ohm's law, since $I_{ij} = (V_i - V_j)Y_{ij}$, and therefore $\bbX$ captures the complete internal state of the electrical grid.

We can similarly describe the \emph{total} complex power demanded and generated at each node, $\bbS^d, \bbS^g \in \reals^{N \times 2}$, as graph signals such that
\begin{subequations}
\begin{align}
    \bbS^d &= \begin{bmatrix} 
        \vdots & \vdots \\
        \sum_{i \in \ccalB^D_i} \Re{S_i^d} & \sum_{i \in \ccalB^D_i} \Im{S_i^d} \\
        \vdots & \vdots
    \end{bmatrix}\\
    \bbS^g &= \begin{bmatrix} 
        \vdots & \vdots\\
        \sum_{i \in \ccalB^G_i} \Re{S_i^g} & \sum_{i \in \ccalB^G_i} \Im{S_i^g}\\
        \vdots & \vdots
    \end{bmatrix}
\end{align}
\end{subequations}
By definition the net power injection at a bus is the total power generated at a bus minus the total power demanded at the bus. Therefore $\bbp$, $\bbq$, $\bbS^g$, and $\bbS^d$ are related by
\begin{equation}\label{eq:state_generation_demand}
    \begin{bmatrix} \bbp & \bbq \end{bmatrix} = \bbS^g - \bbS^d.
\end{equation}

Additionally, we define a cost function ${C: \reals^{N \times 4} \to \reals}$ that maps the state $\bbX$ to a real valued cost. To do this we assume that there is no more than one generator per node, $|\ccalB^G_i| \le 1$. Hence, using \eqref{eq:cost} and \eqref{eq:state_generation_demand}, we can define the cost function as 
\begin{equation}
    C(\bbX) := \sum_{i \in \ccalB} \sum_{j \in \ccalB^D_i} c_j(\bbS^g_i).
\end{equation}
Note that we can approximate grids with multiple generators per bus by forming an equivalent grid with additional auxiliary buses, which are connected by branches with a large admittance.

The solution of the optimal power flow problem is a feasible state which minimizes this vector cost function.
\begin{equation}\label{eq:opf_short}
    \argmin_{\bbX \in \ccalX (\bbS^d)} C(\bbX)
\end{equation}
where $\ccalX(\bbS^d)$ is the feasible set -- the set of states $\bbX$ that satisfy the optimal power flow constraints (see Table \ref{tab:OPF}). Since equation \eqref{eq:power_flow} depends on elements of $\bbS^d$, so does the feasible set.

As explained in \ref{subsec:solutions} finding the solution of \eqref{eq:opf_short} is difficult.  In practice, we are interested minimizing the average cost over time. That is, we want to find a map (likely non-linear) $\bbPhi(\bbS^d)$ that minimizes the expected value of the cost function, given some unknown distribution of power demand, $\bbS^d \sim \ccalD$.
\begin{equation}\label{eq:min_expectation}
    \argmin_{\bbPhi(\bbS^d) \in \ccalX(\bbS^d)} \mathbb{E}_\ccalD \left[C(\bbPhi(\bbS^d)) \right]
\end{equation}

Solving \eqref{eq:min_expectation} in its generality is typically intractable \cite{lan2016algorithms}. Therefore, we choose a parametrization of the map $\bbPhi(\bbS^d) = \bbPhi(\bbS^d ; \ccalH)$ where $\bbPhi$ now becomes a known family of functions (a chosen model) that is parameterized by $\ccalH$. Likewise, to address the issue of the unknown distribution $\ccalD$, we assume the existence of a dataset $\ccalT = \{\bbS^d\}$ that can be used to approximate the expectation operator \cite{lan2016algorithms}, giving rise to the well-studied empirical risk minimization (ERM) problem \cite{Vapnik99-OverviewStatisticalLearning}. With this in place, the problem now boils down to choosing the best set of parameters $\ccalH$ that solve
\begin{equation} \label{eq:objective_parameter}
    \ccalH^* = \argmin_{\ccalH, \bbPhi(\bbS^d; \ccalH) \in \ccalX} \quad \sum_{\bbS^d \in \ccalT} C(\bbPhi(\bbS^d)).
\end{equation}

The desirable properties of locality and scalability can be achieved by a careful choice of the model $\bbPhi(\bbS^d;\ccalH)$. In particular, we focus on graph neural networks (GNNs; refer to section \ref{sec:GNN}) \cite{Bruna14-DeepSpectralNetworks, Defferrard17-CNNGraphs, Gama19-Architectures} to exploit their stability properties \cite{Gama19-Stability} that guarantee scalability \cite{Eisen19-Wireless, Tolstaya19-Flocking}. If we choose the above defined adjacency matrix $\bbA$ as the GSO, the model becomes $\bbPhi(\bbS^d) = \bbPhi(\bbS^d; \ccalH, \bbA)$.

\subsection{Additional architectural considerations}

To naturally enforce the inequality constraints, we can choose the final point-wise non-linearity function $\sigma_L$ such that the generation \eqref{eq:generator_limit} and voltage magnitude constraints \eqref{eq:voltage_magnitude} are always satisfied. In general, consider a constraint of the form $a \le x \le b$ where $x$ is the variable and $a \le b$. We can define,
\begin{equation}
    \gamma(x; a, b) =  \frac{b-a}{1+e^{-x}} + a
\end{equation}
which is a scaled and shifted sigmoid function and it output will always satisfy the constraint. Let $\tilde\bbX$ be the output of the GNN before the final point-wise non-linearity is applied, such that $\bbPhi(\bbS^d; \ccalH, \bbA) = \sigma_L(\tilde\bbX) = \bbX$. We can pick our non-linearity such that $V_i = \gamma(\tilde{V}_i; \tilde{V}_{i,\min}, \tilde{V}_{i,\max})$ and $S^g_{i} = \gamma(\tilde{S}^g_{i}; \tilde{S}^g_{i,\min}, \tilde{S}^g_{i,\max})$. The exact definition of $\bbX$ follows from equations \eqref{eq:state} and \eqref{eq:state_generation_demand}.

Additionally, we can augment the input to include the lower and higher bounds which are different at each bus. More specifically, since the buses in the electrical grid are heterogeneous, the values of the bus constraints such as $S^g_{i,\min}$ and $S^g_{i,\max}$ vary. To express this heterogeneity in a graph neural network we augment the input, $\bbS^d$, with the constraints from table \ref{tab:OPF}. Define $\bbU \in \reals^{N \times 8}$ be the input to the GNN such that
\begin{equation}
    \bbU = \begin{bmatrix}
    \bbS^d & \bbS^g_{\min} & \bbS^g_{\max} & \bbV_{\min} & \bbV_{\max}
    \end{bmatrix}
\end{equation}
where $\bbS^g_{\min}$ and $\bbS^g_{\max}$ are matrices in $\reals^{N \times 2}$ and $\bbV_{\min}$ and $\bbV_{\max}$ are vectors in $\reals^{N}$. With this change, the model becomes $\bbPhi(\bbU; \ccalH, \bbA)$.


\subsection{Optimization using penalty functions}\label{subsec:penalty_method}

Finding a GNN parametrization \eqref{eq:objective_parameter} to the OPF problem is challenging since it is a (non-convex) constrained optimization problem. Instead of solving the problem directly, we can approximate the constraints using \emph{penalty functions}. To illustrate this, consider that the optimal power flow problem can be expressed in standard form,
\begin{subequations}
\begin{align}
    \min\quad& C(\bbX)&\\
    \st\quad& g_i(\bbX) \le 0   & i = 1,...,n \label{eq:standard_inequality}\\
    & h_i(\bbX) = 0             & i = 1,...,m \label{eq:standard_equality}
\end{align}
\end{subequations}
where $g_i$ and $h_i$ are (non-convex) functions that capture all the constraints described in Table \ref{tab:OPF}. The inequality constraints \eqref{eq:generator_limit}, \eqref{eq:voltage_magnitude}, \eqref{eq:power_max}, and \eqref{eq:voltage_magnitude} can be expressed in the form \eqref{eq:standard_inequality}. For example, \eqref{eq:power_max} can be expressed as
$$g_{2i}(\bbX) = S^g_{i,\min} - S^g_i \le 0$$ and $$g_{2i-1}(\bbX) = S^g_i - S^g_{i,\max} \le 0$$
for all $i \in \ccalB^G$. Similarly, equation \eqref{eq:bus_power} can be expressed as 
$$h_i(\bbX) = S_i - Y^s_i|V_i|^2 - \sum_{i,j \in \ccalE \cup \ccalE^T} S_{ij} = 0$$ 
for all $i \in \ccalB$. 

Each inequality constraint $g_i(\bbX) \le 0$ is handled by adding a penalty function $\phi_i(\bbX)$ to the objective. Similarly, the equality constraints are replaced by $\psi_i(\bbX)$. The penalty functions quantify the degree to which the constraints are violated. This allows us to approximate \eqref{eq:objective_parameter} by an unconstrained optimization problem \eqref{eq:minimize_loss}. 

The \emph{loss} function, $\ccalL(\bbX)$, combines the cost function and the penalty functions,
\begin{equation}\label{eq:loss}
    \ccalL(\bbX) = C(\bbX) + \sum_{i=1}^n \lambda_i \phi_i(\bbX) + \sum_{i=1}^m \mu_i \psi_i(\bbX)
\end{equation}
where $\lambda_i, \mu_i \ge 0$ are weight parameters. Instead of solving \eqref{eq:objective_parameter} directly, we instead minimize the loss,
\begin{equation}\label{eq:minimize_loss}
    \hat{\ccalH} = \argmin_{\ccalH} \sum_{\bbS^d \in \ccalT} \ccalL(\bbPhi(\bbS^d; \ccalH, \bbA))
\end{equation}
which is an unconstrained minimization problem. If we chose differentiable $\phi_i,\psi_i$ and assuming that $C(\bbX)$ is also differentiable then we can find the solution to \eqref{eq:minimize_loss} using gradient descent.

Unlike $\ccalH^*$ in \eqref{eq:objective_parameter}, the parameters given by the solution of \eqref{eq:minimize_loss}, $\hat{\ccalH}$, no longer guarantee that $\bbPhi(\bbS^d; \hat{\ccalH}) \in \ccalX(\bbS^d)$ is feasible for all $\bbS^d \in \ccalT$. This is the trade-off we make by approximating the problem with penalty functions: for some inputs our outputs might be infeasible for the constrained problem. Note however that it is inexpensive to check whether our approximate solution is infeasible, and if that is the case we can fall back on slower traditional solvers. Furthermore, by careful choice of penalty functions we can ensure a low rate of constraint violations.


\subsection{Choice of penalty functions}\label{subsec:penalty_choice}

Our choice of the inequality penalty, $\phi_i(\bbX)$, is inspired by the log-barrier method \cite{Boyd04-Convex}. In the log-barrier method a logarithmic function is used instead to approximate the inequality constraints. The log-barrier function has the form $-(1/t)\log(-g_i(\bbX))$ where $t > 0$ is a parameter. Higher values of $t$ provide a fine approximation to the indicator function. Notice that the log-barrier function requires the state to be feasible, $\bbX \in \ccalX(\bbS^d)$, otherwise the logarithm's output is undefined. However, when utilizing gradient descent for GNN training, the initial parameters $\ccalH$ are randomly initialized and do not produce feasible solutions. That is, $g_i(\Phi(\bbS^d; \ccalH, \bbA)) < 0$ for many $\bbS^d \in \ccalT$.

\begin{figure}
    \centering
    \begin{tikzpicture}
\begin{axis}[
    ticks=none,
    legend pos=north west, 
    axis x line=center, 
    axis y line=center,
    xlabel={$u$},
    height=5cm,
    width=10cm,
]
    
    \def\s{2}
    \def\t{1}
    \addplot[domain=-(\t/\s):0.5, color=blue, thick]{-\s*(-x-\t/\s) - \t*ln(\t/\s)};
    \addlegendentry{$-\overline{\log}_s(-u)$};
    
    \addplot[domain=-(\t/\s):-0.01,color=red, thick]{-\t*ln(-x)};
    \addlegendentry{$-\log(-u)$};
    
    \addplot[domain=-2:-(\t/\s), thick]{-\t*ln(-x)};
    \addlegendentry{both}
    
    \filldraw[black] (0,0) circle[radius=2pt] node[anchor=south east]{(0,0)};
\end{axis}
\end{tikzpicture}
    \caption{Comparison of the log-barrier (red) and extended log-barrier (blue) functions. The later uses the extended logarithm \eqref{eq:extended_log} instead of the logarithm. As a result, the extended log-barrier follows the traditional log-barrier until the slope reaches $s$; from there, it continues linearly.}
    \label{fig:log_barrier}
\end{figure}
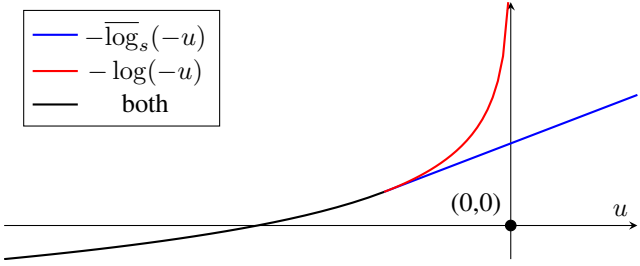

To overcome this, we define the extended logarithm, $\overline{\log}_s$, which is a linear piece-wise extension of the logarithm to $\reals_-$. We will use this extended logarithm to define the extended log-barrier function. The parameter $s \in \reals$ defines the maximum value of the function's derivative.
\begin{subequations}\label{eq:extended_log}
\begin{align}
    \overline{\log}_s(u) &:= 
    \begin{cases}
    \log(u) & \text{if } (u \ge 1/s)\\
    s(u + \frac{1}{s}) - \log(\frac{1}{s}) & \text{otherwise}
    \end{cases}\\
    \tfrac{d}{du}\overline{\log}_s(u) &:= \begin{cases} 
    \min(1/u, s) & \text{if } (u > 0)\\
    s & \text{otherwise}
    \end{cases}
\end{align}
\end{subequations}
We define the extended log-barrier as $-(1/t)\overline{\log}_s(-u)$. Unlike the traditional log-barrier it is defined on $\reals_+$ and therefore provides a \emph{soft} barrier (see figure \ref{fig:log_barrier}).

Hence, we define the inequality penalty function to be the extended log-barrier function,
\begin{equation}\label{eq:inequality_penalty}
    \phi_i(\bbX) := -(1/t)\overline\log_s(-g_i(\bbX))
\end{equation}
where $s \ge 0$ is a parameter. We omit the scaling parameter $t$, which is lumped into $\lambda_i$ instead. In practice we want $s$ to be as large as possible without leading to numerical overflow. Meanwhile, we use the square function as the penalty for the equality constraint \eqref{eq:standard_equality},
\begin{equation}
    \psi_i(\bbX) := h_i(\bbX)^2.
\end{equation}
Note that both penalty functions are differentiable, which is necessary to run gradient descent.

\section{Numerical Experiments} \label{sec:sims}



\iflong
\def\figwidth{1.0}
\else
\def\figwidth{0.7}
\fi

To evaluate the efficacy of our proposed architecture we run simulations on the IEEE-30 and IEEE-118 power system test case networks. 
\iflong Figure~\ref{fig:118} shows a visualization of the IEEE-118 test case topology.\fi
The 30 bus test case is used to run a hyper-parameter grid search. The best set of hyper-parameters is used to train models for the 30 and 118 bus systems, which are compared against an interior point method solution. For each model we compare the dollar cost for electricity generation and the rate of constraint violations.

\iflong
\begin{figure}
	\centering
	\includegraphics[width=0.9\linewidth]{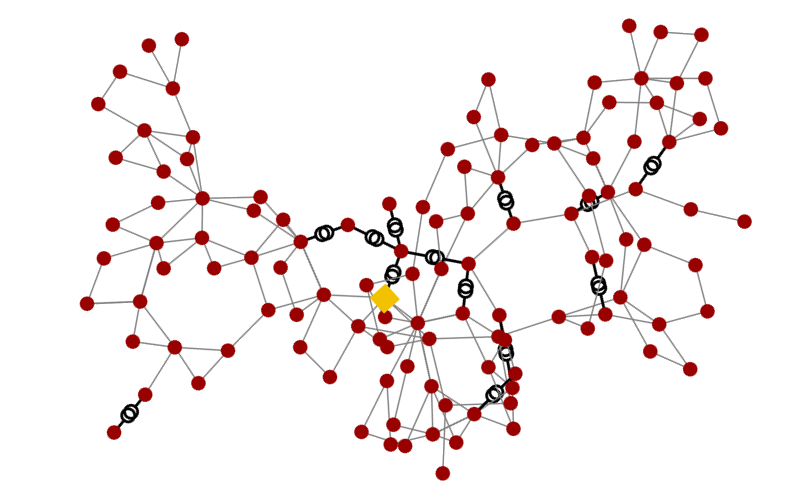}
	\caption{Diagram of IEEE-118 power system test case. Red circles are buses, the yellow square is an external grid connection and intersecting circles are transformers.\vspace{-0.3cm}}
	\label{fig:118}
\end{figure}
\fi

In section \ref{subsec:dataset_generation} we explain how we generate synthetic datasets for training and testing the GNN. Then in section \ref{subsec:error} we define the \textit{constraint violation rate}, a metric that measures how often GNN solutions are infeasible. Together, generation cost and constraint violation rate, help us quantify the performance and robustness of solutions. Specifically, in section \ref{subsec:search} we perform a hyper-parameter search on the IEEE-30 test case. The hyper-parameters chosen from the IEEE-30 test case are also used to train the IEEE-118 model. Finally, in the remaining sections we discuss in depth the performance of the models trained on the IEEE-30 and IEEE-118 test cases.

\subsection{Dataset Generation} \label{subsec:dataset_generation}

We construct datasets based on IEEE power system test cases provided by MATPOWER \cite{matpower}. The test cases provide reference values for complex power demanded by each load $S^d_{i,\text{ref}} \in \mathbb{C}$ for all $i \in L$.
\begin{align*}
    \Re{S^d_{i,\text{ref}}} &= \sum_{i \in L_i} \Re{S_{i,\text{ref}}^d} \\
    \Im{S^d_{i,\text{ref}}} &= \sum_{i \in L_i} \Im{S_{i,\text{ref}}^d}  
\end{align*}
for all $i \in \ccalB$.
Following \eqref{eq:state_generation_demand}, let $\bbS^d_{\text{ref}} \in \mathbb{R}^{N \times 2}$ be total reference power demanded at each node such that,
\begin{equation}
    \bbS^d_{\text{ref}} = \begin{bmatrix} 
    \vdots & \vdots\\
    \Re{S^d_{i,\text{ref}}} & \Im{S^d_{i,\text{ref}}}\\
    \vdots & \vdots
    \end{bmatrix}
\end{equation}
We obtain a dataset by sampling (element-wise) from a uniform distribution around the reference load, following the same methodology used in \cite{guha},
\begin{equation} \\
\bbS^d \sim \text{Uniform}(0.9 \bbS^d_{\text{ref}},\; 1.1 S^d_{\text{ref}}) \label{eq:load_distribution}.
\end{equation}
Therefore, for both IEEE-30 and IEEE-118 test cases, we generate $100,000$ training samples and $1,000$ test samples. For each sample from the test set we solve the corresponding OPF problem using IPOPT \cite{ipopt, pandapower} to obtain ground truth solutions $\bbX^*$. These are \emph{not} used in training. Instead they serve as a benchmark to compare performance of the GNN against solving OPF directly. Note that IPOPT does not converge for all $1,000$ test samples, and thus we discard the samples in the test set for which we cannot get a baseline.

\subsection{Measuring Constraint Violations}\label{subsec:error}

In our experiments we evaluate model both in terms of constraint violation rate and severity. However, OPF constraints have different units and magnitudes. Therefore to rigorously define both, we normalize the constraints.

\def\xmin{x_{\text{min}}}
\def\xmax{x_{\text{max}}}

Consider a constraint of the form $\xmin \le x \le \xmax$ where $x \in \mathbb{R}$ is a decision variable.
We define the absolute error as
\begin{equation}\label{eq:abs_error}
   E(x) = (x-\xmax)^+ + (x-\xmin)^-
\end{equation}
When $\xmin \ne \xmax$ we could define the relative error as 
\begin{equation}\label{eq:relative_error_naive}
    E(\xmin, x, \xmax) / (\xmax-\xmin),
\end{equation}
so that relative error measures the relative deviation from the feasible region. However, this is undefined whenever $\xmin = \xmax$, so in those cases we normalize based on the average feasible set size of constraints of the same type. For instance, the average over all power generation constraints \eqref{eq:generator_limit}. Specifically let us rewrite the inequality constraints from \eqref{eq:generator_limit}, \eqref{eq:voltage_magnitude}, \eqref{eq:power_max}, and \eqref{eq:voltage_magnitude} as
\begin{equation}
    x_{ij,\text{min}} \le g_{ij}(\bbX) \le x_{ij,\text{min}}
\end{equation}
where $g_{ij}$ is a function from the state matrix to the reals. Each value of $i$ represents a different constraint type, with the real and imaginary components of the aforementioned inequality constraints having distinct indices. Hence, $j \in \{1,...,N_i\}$ is an index over the buses, generators or edges, depending on the constraint type so that $N_i \in \{|\ccalB|, |\ccalB^G|, |\ccalE| \}$.

Denote the feasible region size from \eqref{eq:relative_error_naive} as 
$$f_{ij} = |x_{ij,\max} - x_{ij,\min}|.$$
This allows us to define the relative error
\begin{equation}\label{eq:error}
    \hat{E}(g_{ij}(\bbX)) = E(g_{ij}(\bbX)) / \eta_{ij} \\
\end{equation}
where
\begin{equation}
    \eta_{ij} = \begin{cases}
        f_{ij} &\text{if } f_{ij} > 0 \\
        \sum_{k \in \{ k \mid f_{kj} > 0 \}} \frac{f_{kj}}{|\{k \mid f_{kj} > 0\}|} &\text{if } f_{ij} = 0
    \end{cases}
\end{equation}

We say a constraint violation occurs whenever \mbox{$E(g_{ij}(\bbX))>0$}. Therefore the constraint violation rate, $R(\bbX)$ is given by
\begin{equation}\label{eq:violation_rate}
    R(\bbX) = \left( \prod_i N_j \right)^{-1} \sum_i \sum_{j \in \{1,...,N_i\}} \mathbb{I}[E(g_{ij}(\bbX)) > 0] 
\end{equation}
where $\mathbb{I}$ is the indicator function. And $i$ sums over all the inequalities from table \ref{tab:OPF}.

\subsection{Hyper-parameter Search}\label{subsec:search}

We picked the parameters based on a hyper-parameter search, during which we tested $486$ combinations of parameter values. For each combination, we train a GNN model for 100 epochs with a batch size of 2048 using the method from section \ref{sec:unsupervisedOPF}. 

We determine the best hyper-parameters by looking at the generation cost \eqref{eq:cost} and constraint violations of the corresponding model on the test set. None of the trained models were able to satisfy all the constraints in table \ref{tab:OPF} for every load distribution in the test set. Therefore, we picked the model with the lowest generation cost out of those which had a constraint violation rate of less than 1\%. The hyper-parameters of this model are shown in table \ref{tab:search_values}.

\begin{table}
    \centering
    \caption{A model was trained for every combination of search values. The chosen value column shows which hyper-parameter values produced the best result.}
    \begin{tabular}{ccc}
        \toprule
        Parameter & Search Values & Chosen Value \\
        \midrule
        K & 2, 4, 8 & 8\\
        F & 16, 32, 64 & 32\\
        s & 10, 100, 500 & 10\\
        t & 10, 100, 500 & 500\\
        L & 1, 2 & 2 \\
        $\eta$ & 1e-5, 1e-4, 1e-3 & 1e-4\\
        \bottomrule
    \end{tabular}
    \label{tab:search_values}
\end{table}

\begin{table}[ht]
    \centering
    \caption{The effect of $F$ and $K$ on average generation cost and constraint violation rate over the test set. The selected parameters are in bold.}
    \begin{tabular}{lrrrrrr}
\toprule
 & \multicolumn{3}{r}{Generation Cost} & \multicolumn{3}{r}{Constraint Violation Rate} \\
F & 16 & 32 & 64 & 16 & 32 & 64 \\
K &  &  &  &  &  &  \\
\midrule
2 & 4.5939 & 1.3213 & 1.1628 & 0.0906 & 0.0441 & 0.0765 \\
4 & 3.8341 & 1.1534 & 1.2134 & 0.0878 & 0.0722 & 0.0871 \\
8 & 1.3823 & \bfseries 1.1905 & 1.1565 & 0.0060 & \bfseries 0.0093 & 0.0510 \\
\bottomrule
\end{tabular}

    \label{tab:search_FK}
\end{table}

Table~\ref{tab:search_FK} shows the effect of varying $F$ and $K$ on the generation cost and violation rate. We vary $F$ and $K$ while keeping the other hyper-parameter values fixed to the chosen values in table \ref{tab:search_values}. Similarly table \ref{tab:search_ts} shows the effect of $t$ and $s$, while table \ref{tab:search_etaL} shows the effect of $\eta$ and $L$. 

We note that for low values of features $(F=16)$ it is required to set a larger value of $K$ to achieve enough expressive power. Even for larger amounts of features such as $F=32$ and $F=64$, increasing the number of filter taps improves the constraint violation rate. Since we are performing the parameter with a 30 node graph, it is likely that larger numbers of filter taps could be more effective on graphs with a larger diameter.

\begin{table}[ht]
    \centering
    \caption{The effect of $t$ and $s$ on average generation cost and constraint violation rate over the test set. The selected parameters are in bold.}
    \begin{tabular}{lrrrrrr}
\toprule
 & \multicolumn{3}{r}{Generation Cost} & \multicolumn{3}{r}{Constraint Violation Rate} \\
t & 10 & 100 & 500 & 10 & 100 & 500 \\
s &  &  &  &  &  &  \\
\midrule
10 & 1.1512 & 1.1673 & \bfseries 1.1905 & 0.0394 & 0.0685 & \bfseries 0.0093 \\
100 & 1.3631 & 1.3474 & 1.4386 & 0.0262 & 0.0028 & 0.0189 \\
500 & 1.3263 & 1.2807 & 1.5720 & 0.0334 & 0.0384 & 0.0686 \\
\bottomrule
\end{tabular}

    \label{tab:search_ts}
\end{table}

Similarly, table~\ref{tab:search_ts} shows the effects of the log-barrier function parameters, as defined in \eqref{eq:inequality_penalty}, on generation cost and violation rate. Here the results are counter intuitive. One would expect that as $t$ rises then the generation cost decreases as the magnitude of inequality penalty function \eqref{eq:inequality_penalty} is inversely proportional to $t$. The opposite occurs, higher values of $t$ trade-off higher generation cost for fewer constraint violations. Similarly, $s$, affects the metrics surprisingly. Higher values of $s$ increase the magnitude of the gradient of $\nabla_\bbX \phi_i$, which should penalize violations more heavily. Therefore, we would expect that higher values of $s$ would reduce the violation rate. While such values do increase the generation cost, they simultaneously increase the violation rate. Perhaps what is happening is numerical instability due to high values of $s$. A critical feature of log-barrier methods is increasing the value of $t$ with each iteration. At the very least, these results indicate that more work could be done to either fine tune $s$ and $t$ or to adapt them during training.

\begin{table}[ht]
    \centering
    \caption{The effect of $\eta$ and $L$ on average generation cost and constraint violation rate over the test set. The selected parameters are in bold.}
    \begin{tabular}{lrrrrrr}
\toprule
 & \multicolumn{3}{r}{Generation Cost} & \multicolumn{3}{r}{Constraint Violation Rate} \\
$\eta$ & 1e-5 & 1e-4 & 1e-3 & 1e-5 & 1e-4 & 1e-3 \\
L &  &  &  &  &  &  \\
\midrule
1 & 4.2703 & 3.9904 & 4.0481 & 0.0880 & 0.0878 & 0.0878 \\
2 & 4.8803 & \bfseries 1.1905 & 1.1776 & 0.0933 & \bfseries 0.0093 & 0.0462 \\
\bottomrule
\end{tabular}

    \label{tab:search_etaL}
\end{table}

 Finally, table~\ref{tab:search_etaL} shows the impact of the step size, $\eta$, and the number of layers, $L$, on generation cost and constraint violation rate. For $L = 1$ the step size has less impact on the model performance than for $L=2$, though in general the generation cost decreases with step size. For $L=2$ this effect is more pronounced. However, both a low and high value $\eta$ cause an increase in violation rate. This is in line with discussion above, with regards to the numerical instability of the inequality constraints.

\subsection{IEEE-30}\label{subsec:ieee30}

Using the parameters determined in section \ref{subsec:search} we train a model for longer: using more epochs and a smaller batch size. Specifically we train a model with $K = 8, F = 32, s = 10, t = 500, L = 2$ and $\eta=1 \times 10^{-4}$. The batch size was $256$ and the model was trained for $1000$ epochs and validation was done after every epoch. In the rest of the section we consider the parameters which minimized the validation loss.

\begin{figure}[ht]
     \centering
     \includegraphics[width=\figwidth\linewidth, trim={0 3mm 0 3mm}, clip]{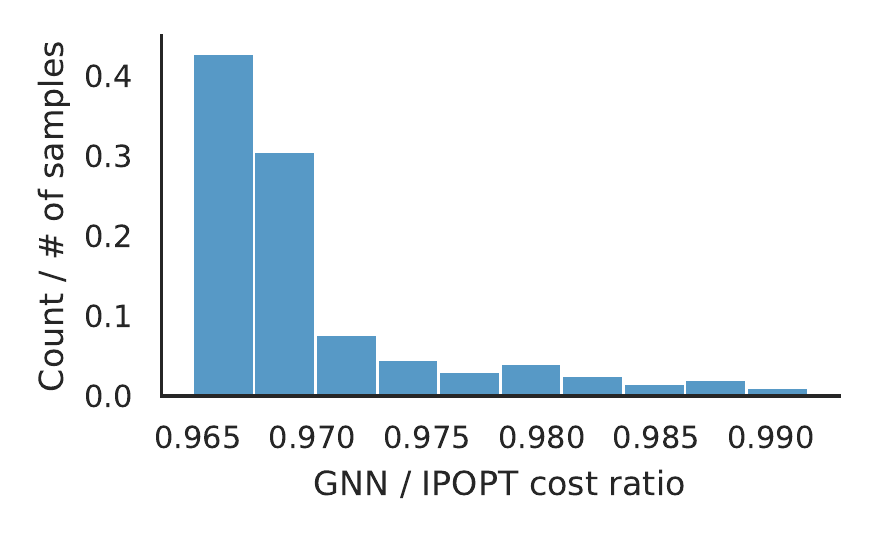}
     \caption{The distribution of GNN cost as a fraction of IPOPT cost for test samples where the GNN did not make any constraint violations on the IEEE-118 dataset.}
     \label{fig:case30/costs}
 \end{figure}

After training, the GNN achieves an average cost of $4.142$. Meanwhile the IPOPT cost is $2.680\%$ greater, at $4.253$. The GNN has an average violation rate of $1.276\%$, as defined by \eqref{eq:violation_rate}, over the test set and at least one violation occurs in 75.44\% of the test samples. As figure~\ref{fig:case30/costs} shows the ratio of the GNN cost to the IPOPT cost for the test samples where \emph{no constraint violations} were made. Notice that the GNN has a lower cost than the IPOPT solution for all test cases. This shows that a GNN solution can consistently find a better feasible solution than the IPOPT method.

\begin{figure}[ht]
    \centering
    \includegraphics[width=\figwidth\linewidth, trim={0 3mm 0 3mm}, clip]{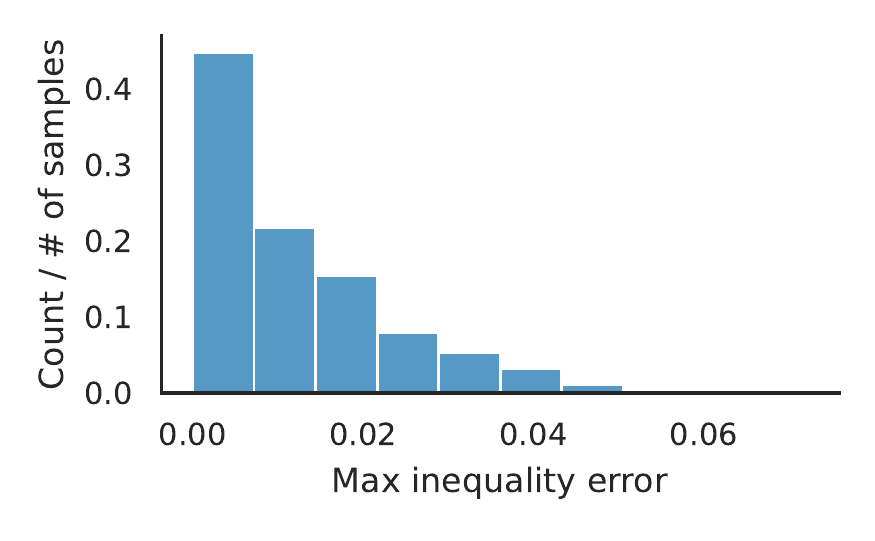}
    \caption{The distribution of the maximum constraint violations of each IEEE-30 test sample.}
    \label{fig:case30/error_max}
\end{figure}

However since there is a violation in at least $75.44\%$ of the test samples, we examine the severity of those violations. Hence, we can look at the distribution of the size of constraint violations as defined by \eqref{eq:error}. Note that all equality constraints are satisfied to within numerical tolerance, therefore we consider only inequality constraint errors. Figure \ref{fig:case30/error_max} shows the distribution of the maximum constraint relative error \eqref{eq:error} size on each test sample. The error is always below $5\%$ and higher errors are less likely. This suggests that violations are due to the shape of the loss function, rather that a failure of the GNN to converge. With more fine tuning of the penalty functions, specifically the $s, t$ parameters, violations could be further minimized.

\begin{figure}[ht]
    \centering
    \includegraphics[width=\figwidth\linewidth, trim={0 3mm 0 3mm}, clip]{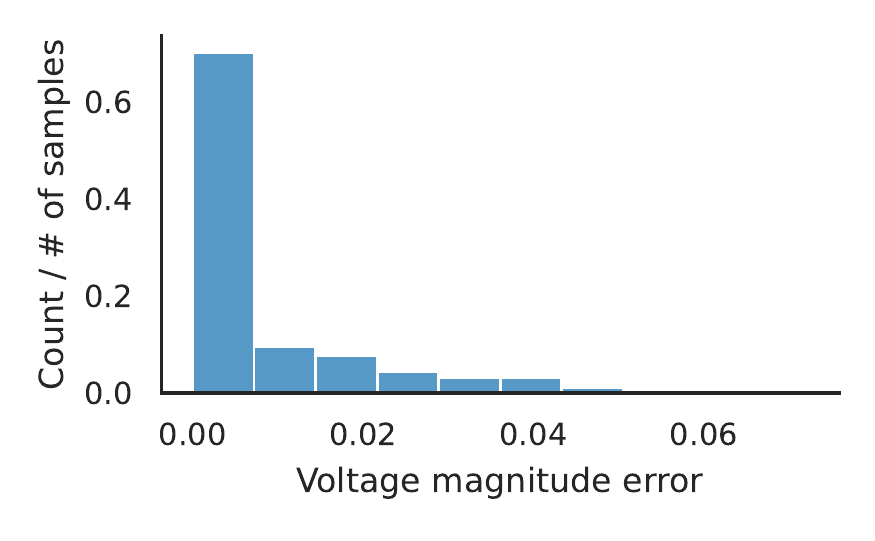}
    \caption{Distribution of the maximum voltage magnitude constraint \eqref{eq:voltage_magnitude} violations of each IEEE-30 test sample.}
    \label{fig:case30/error_vm}
\end{figure}

\begin{figure}[ht]
    \centering
    \includegraphics[width=\figwidth\linewidth, trim={0 3mm 0 3mm}, clip]{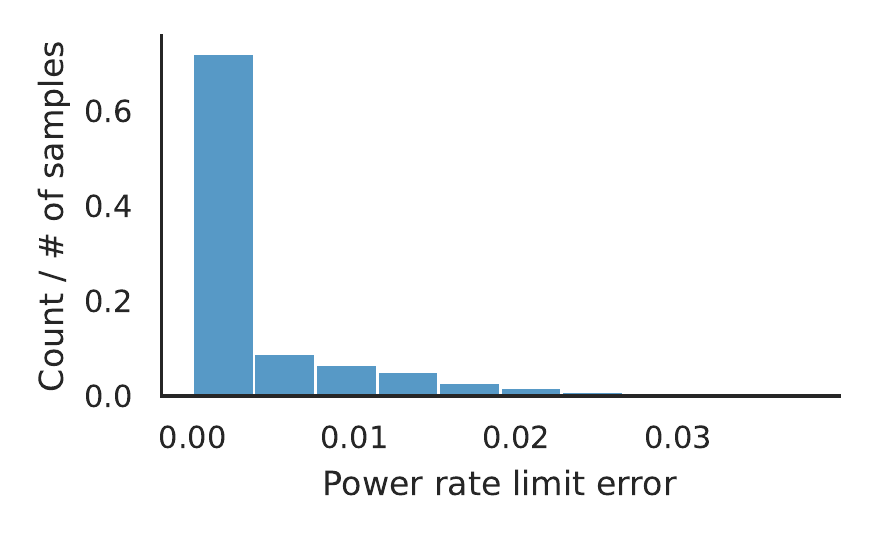}
    \caption{Distribution of the maximum power rate limit \eqref{eq:power_max} violations of each IEEE-30 test sample.}
    \label{fig:case30/error_rate}
\end{figure}

Additionally, we look at the error distributions for individual constraints. Figures \ref{fig:case30/error_vm} and \ref{fig:case30/error_rate} show histograms for inequalities \eqref{eq:voltage_magnitude} and \eqref{eq:power_max}, respectively. The maximum violation for generated power \eqref{eq:power_max} is less than $0.3\%$ and the distribution tail is too small to show on a histogram. We do not show histograms for the remaining constraints, since they are not violated.

Note that voltage magnitude relative errors do not exceed $5\%$. Keep in mind that the feasible region is typically 0.95 to 1.05 per unit, but may be as narrow as 0.99 to 1.01 per unit. In the first case a $5\%$ deviation represents an excess of $0.005$ per unit and in the latter case $0.001$ per unit. From the perspective of power grid operation, this is not a significant deviation if it occurs for a short period of time or is intermittent. 

Similarly, the power rate constraint violations are not severe. In practice there are typically three sets of rate constraints: for long term, short term, and emergency power rates. It is not particularly important that these constraints are satisfied at every time step, but rather in expectation with some limit on the variance of violations. These results suggest that a GNN is a viable alternative to IPOPT for power grid control.

\subsection{IEEE-118}\label{subsec:ieee118}

\begin{figure}[ht]
    \centering
    \includegraphics[width=\figwidth\linewidth, trim={0 3mm 0 3mm}, clip]{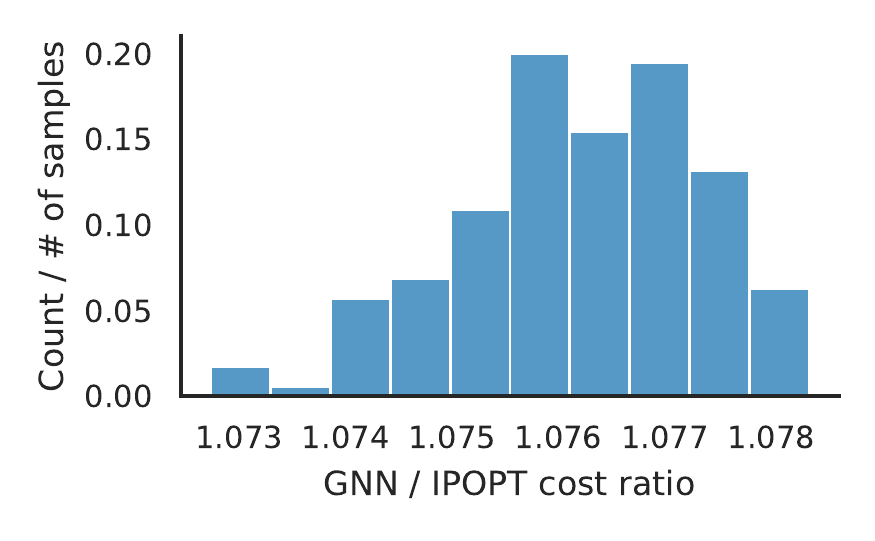}
    \caption{Distribution of cost improvement of the GNN over the IPOPT solution for samples where the GNN did not make any constraint violations on the IEEE-118 dataset.}
    \label{fig:case118/costs}
\end{figure}

\begin{figure}[ht]
    \centering
    \includegraphics[width=\figwidth\linewidth, trim={0 3mm 0 3mm}, clip]{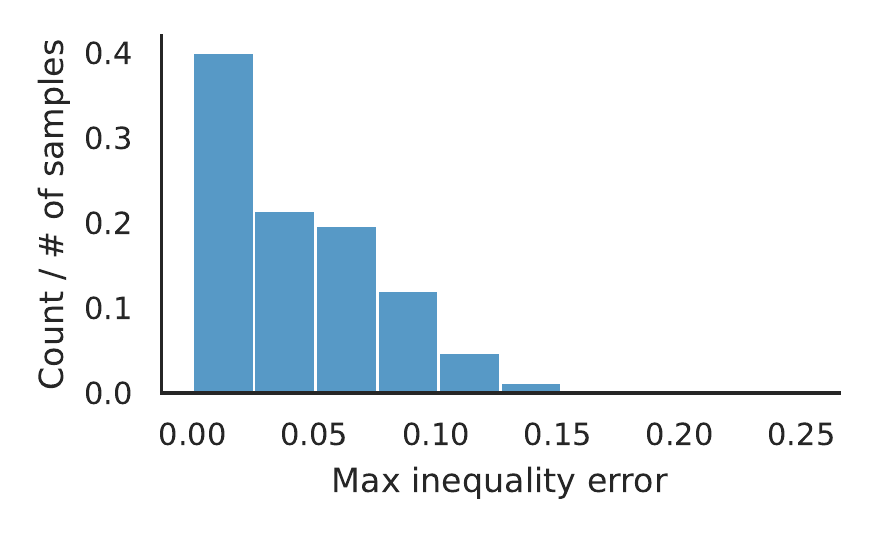}
    \caption{Distribution of the maximum constraint violations on each sample in the IEEE-118 dataset.}
    \label{fig:case118/error_max}
\end{figure}

We trained another model with the same parameters on the IEEE-118 dataset. The GNN achieves an average cost of $903.96$ compared to $836.10$ with the IPOPT solver. The GNN solutions are $8.12\%$ higher than the IPOPT ones. Figure \ref{fig:case118/costs} shows the distribution of the solution cost on the portion of the test set where there were no constraint violations. Note that an $8\%$ higher cost is reasonably close to the IPOPT solution, and as we will explain below, there is an architectural reason why this is unsurprising. 

Constraint violations occur in $82.5\%$ of the test samples. Figure \ref{fig:case118/error_max} shows the distribution of the maximum violation error of each sample on the test set. The constraint violations are much higher than for the IEEE-30 case. However, the tail of the distribution remains short, which shows promise for improvement with further tuning; especially considering that the hyper-parameter search was conducted on the IEEE-30 network.

\begin{figure}[ht]
    \centering
    \includegraphics[width=\figwidth\linewidth, trim={0 3mm 0 3mm}, clip]{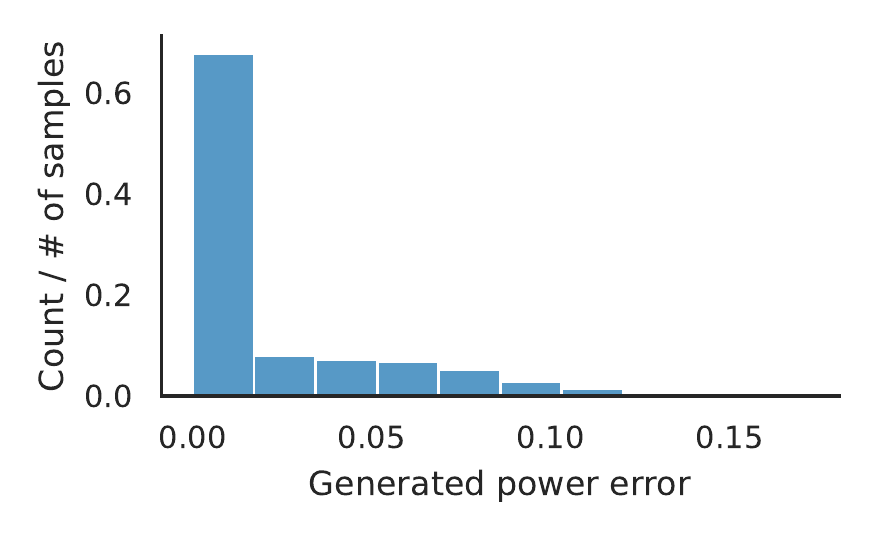}
    \caption{Distribution of maximum generated power \eqref{eq:generator_limit} violations of each IEEE-118 test sample.}
    \label{fig:case118/error_gen}
\end{figure}

\begin{figure}[ht]
    \centering
    \includegraphics[width=\figwidth\linewidth, trim={0 3mm 0 3mm}, clip]{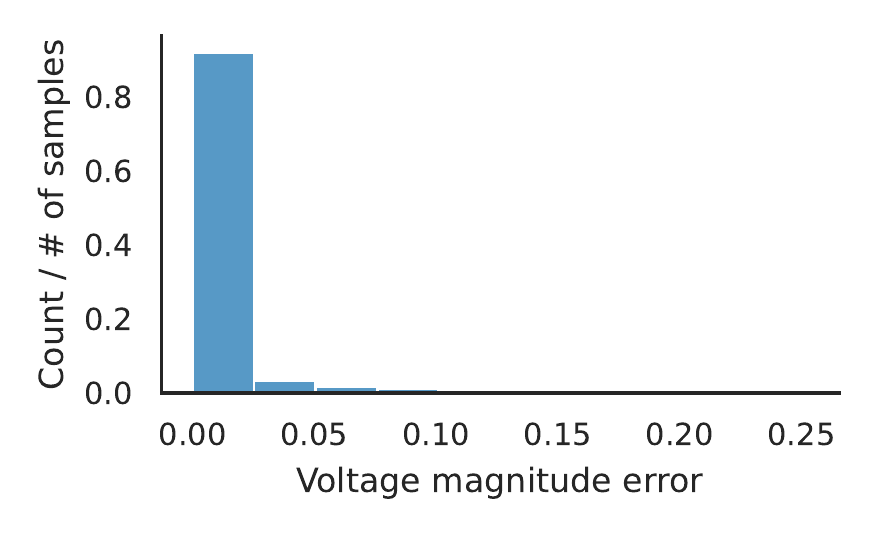}
    \caption{Distribution of maximum voltage magnitude \eqref{eq:voltage_magnitude} violations of each IEEE-118 test sample.}
    \label{fig:case118/error_vm}
\end{figure}

The GNN only violates the generated limit \eqref{eq:generator_limit} and voltage magnitude \eqref{eq:voltage_magnitude} constraints. Figures \ref{fig:case118/error_gen} and \ref{fig:case118/error_vm} show their respective error distributions. Generator limit constraint violations are more common than the voltage magnitude violations. The largest violation has a relative error of $17.20\%$. Meanwhile, while Voltage magnitude constraint violations are rarer, the largest one had a relative error of $25.28\%$. However such high violations are uncommon as can be seen from the histogram.

While the GNN came close to the IPOPT results, there is a clear difference in performance on the IEEE-30 and IEEE-118 datasets. However, it appears that the IEEE-118 dataset violates key assumptions of the GNN model. Note that in the IEEE-118 dataset power rate constraints are not binding. In fact the IEEE-118 test cases have current constraints so high that the would never be binding \cite{Lipka2017-CurrentConstraints}. One key assumption behind the use of a GNN is that there should be a degree of locality to the data. However, since there are no current constraints, then global knowledge is needed to find an optimal solution.


\section{Conclusions} \label{sec:conclusions}



In recent years there has been a renewed interest in the OPF problem. It is critical to the efficient operation of electrical grids, especially as these will have to adapt in the face of climate change. Solving OPF traditionally uses interior point methods, which is computationally costly and does not scale to large networks. Recently, new ways of solving this problem were proposed using modern convex optimization techniques and machine learning.

In this paper we proposed a novel approach to solve the OPF by augmenting the output in order to make penalty functions on constraints differentiable. In addition, we put forward a new way of evaluating model performance on these datasets, beyond cost and violation rate: relative violation error. Our experiments show that a GNN can outperform interior method in the right conditions. Specifically, when current or power rate limits are binding. This is the case with most real power systems where transmission capacity is limited.

We believe this initial work can potentially spark interest in the use of machine learning techniques and, in particular GNNs, to the toolbox of OPF solutions. This may require revisiting the specification of power system test cases and the creation of benchmark datasets and baselines for this problem. We are confident it will invite further exploration in terms of other techniques that can complement GNNs such as sparse operations and scalability.






\bibliographystyle{bibFiles/IEEEtranD}
\bibliography{bibFiles/myIEEEabrv,bibFiles/biblioOPF}

\end{document}